\documentclass[12pt]{iopart}

\usepackage{graphicx}

\begin{document}

\title{Non-trivial length dependence of the conductance and
negative differential resistance in atomic molecular wires}

\author{V M Garc\'{\i}a-Su\'arez and C J Lambert}
\address{Department of Physics, Lancaster University, Lancaster,
LA1 4YB, U. K.}

\ead{v.garcia-suarez@lancaster.ac.uk}

\date{\today}

\begin{abstract}
We study the electronic and transport properties of two novel
molecular wires made of atomic chains of carbon atoms (polyynes)
capped with either, benzene-thiols or pyridines. While both
molecules are structurally similar, the electrical conductance of
benzene-thiol-capped chains attached to gold electrodes is found
to be much higher than that of pyridine-capped chains. We predict
that the conductance is almost independent of molecular length,
which suggests that these molecules could be ideal molecular wires
for sub-10 nm circuitry. Both systems exhibit negative
differential resistance (NDR) but its origin and characteristics
depend on the type of molecule. We find a novel type of NDR
mechanism produced by the movement of the LUMO resonance with
bias. We also show that by gating the pyridine-capped molecules it
is possible to make the NDR disappear and dramatically modify the
$I$-$V$ characteristics and the length dependence.
\end{abstract}


\maketitle

Molecular wires are currently receiving increasing attention due
to recent experimental \cite{Joa95,Ree97,Dat97,Smi02} and
theoretical \cite{Xue02,Bra02,Fuj03,Cal04,Roc06} advances. These
systems could be candidates for substituting silicon components in
nanoscale circuits once the atomic limit is approached. However,
progress in this field has been hindered by the extremely low
conductance found in the vast majority of molecules. This limiting
property is a consequence of the location of the injection energy,
defined by the Fermi level $E_\mathrm{F}$ of the leads, which
typically lies in the HOMO-LUMO (HL) gap. Such a level
misalignment moves the electronic transport to the tunneling
regime, where the current decreases exponentially as the length of
the molecule increases, and makes it very difficult to use
molecular wires as interconnects. For this reason the search for
molecular wires whose resonances are broad enough to reach the
Fermi energy or whose level misalignment is small for any
molecular length is of paramount importance for this field.
Recently polyyne-based molecular wires, made with carbon chains
with an even number of atoms, contacted directly to gold leads by
sulphur atoms \cite{Crl07} were predicted to have a high
length-independent conductance and ohmic behavior for a large
range of biases. Unfortunately these molecules can not be made in
large quantities and contacted to leads because they are extremely
reactive \cite{MBryce}. One possible way of overcoming this
crucial deficiency is by capping the carbon chains with aromatic
rings, which makes them much more stable and easier to manipulate
\cite{MBryce}. The result is a new type of molecules made of a
very high conducting backbone, the polyyne, connected to two less
conducting rings.

\begin{figure}
\includegraphics[width=\columnwidth]{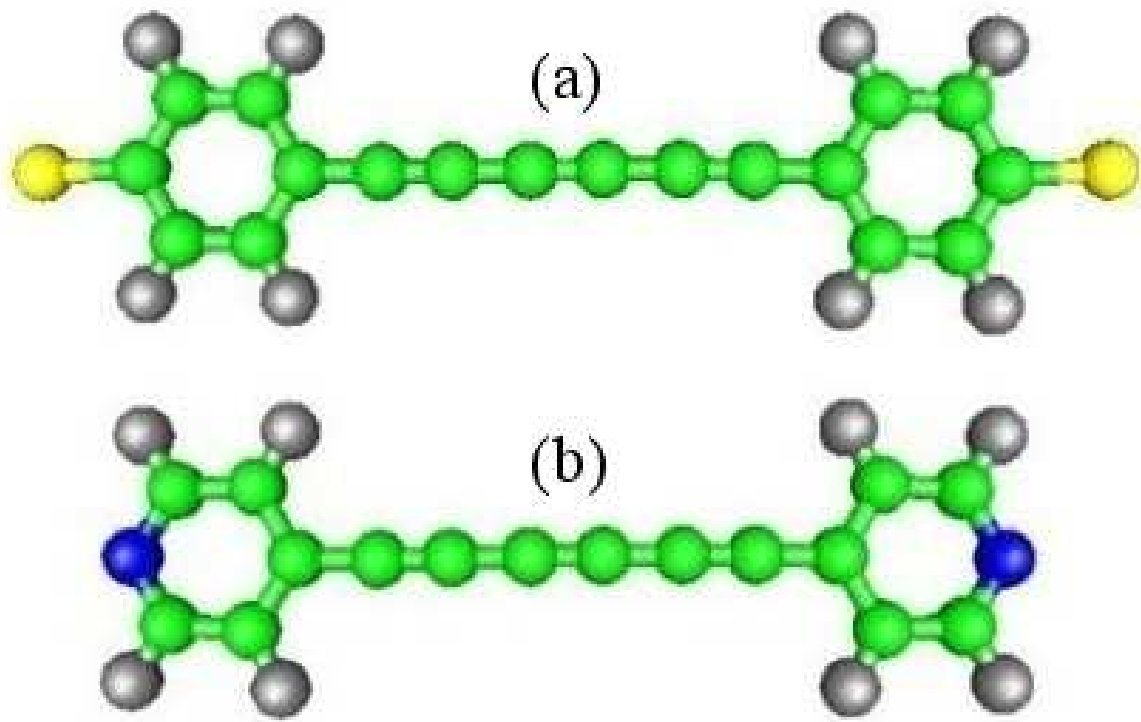}
\caption{\label{Fig1} Polyyne-based molecular wires. (a) Polyyne
between benzene-thiol groups, B-$n$-B, and (b) polyyne between
pyridine groups, P-$n$-P, for $n=3$.}
\end{figure}

In this article we study the electronic and transport properties
of the above mentioned polyynes capped with benzene-thiol
(B-$n$-B) or pyridine groups (P-$n$-P) for $n$ between 1 and 6
(between 2 and 12 atoms in the carbon chain). An example can be
seen in Fig. (\ref{Fig1}) for $n=3$. We predict that this new
family of molecular wires not only possesses an almost
length-independent conductance, but also exhibits negative
differential resistance (NDR), which results in a reduction of the
current as the bias voltage increases. NDR is very important in
the field of electronic technology and also in basic research.
Since its discovery \cite{Esa58} lots of applications in the area
of semiconductor physics have been found, which include
amplification \cite{Mcw66}, digital applications
\cite{Bro98,Mat99}, and oscillators \cite{Bro91}. In the context
of molecular electronics there are various types of mechanisms
that can lead to the phenomenon of NDR. One is due to the movement
of resonances to the bulk silicon gap as the bias is applied
\cite{Rak05}, which allows the possibility of tuning the NDR peak
by varying the coupling between the STM tip and the molecule.
Others include chemical changes \cite{Che99}, the destruction of
conductance resonances as a consequence of the misalignment of
localized or interface states \cite{Inw89,Dal06} and local orbital
symmetry matching \cite{Che07}. Here we show the presence of two
types of NDR, which depend on the type of end-group that cap the
polyyne. The first one, which is found in the B-$n$-B molecules,
is due to the destruction of the HOMO resonance by the
bias-induced asymmetry; the second one, which is found in the
P-$n$-P molecules, is a novel class of NDR produced by the
bias-induced movement of the LUMO resonance.

To obtain these predictions we used density functional theory
(DFT) \cite{Koh65} as implemented in the SIESTA code \cite{Sol02},
which employs norm-conserving pseudopotentials to get rid of the
core electrons and linear combinations of atomic orbitals to
expand the valence states. The basis set included double zetas and
polarization orbitals (DZP) both in the simulations to perform the
structural relaxations and the transport calculations. We used the
local density approximation (LDA) to compute the exchange and
correlation energy, which works rather well for light elements and
systems where electrons are delocalized. The Hamiltonian, overlaps
and electronic densities were evaluated in a real space grid
defined with a plane wave cutoff of 200 Ry. The dimensions of the
unit cell were long enough along $x$ and $y$ to avoid overlaps and
strong electrostatic interactions with molecular images. The
molecular coordinates were relaxed until all forces were smaller
than 0.05 eV/\AA\/. To obtain the transport properties we used the
SMEAGOL code \cite{Roc06}, which is interfaced to SIESTA and uses
the non-equilibrium Green's functions formalism (NEGF) to
calculate self-consistently the charge density, the transmission
coefficients and the $I$-$V$ characteristics. The system was
divided in three parts: a left lead, a right lead and an extended
molecule (molecule + part of the leads modified by the presence of
the molecule and the surfaces) and a separate calculation was
performed for each of them \cite{LeadsCalc}. In the bulk
calculation the unit cell was made of three slices of fcc gold
grown along (111), with 9 atoms per slice and periodic boundary
conditions along $x$, $y$ and $z$, but only $k$-points along $z$,
which we chose as the transport direction. In the transport
calculation we had to include one of the previous unit cells on
each side plus two slices of the same material on the left and
right parts of the extended molecule, respectively. Each transport
calculation involved around 120 atoms.

\begin{figure}
\includegraphics[width=\columnwidth]{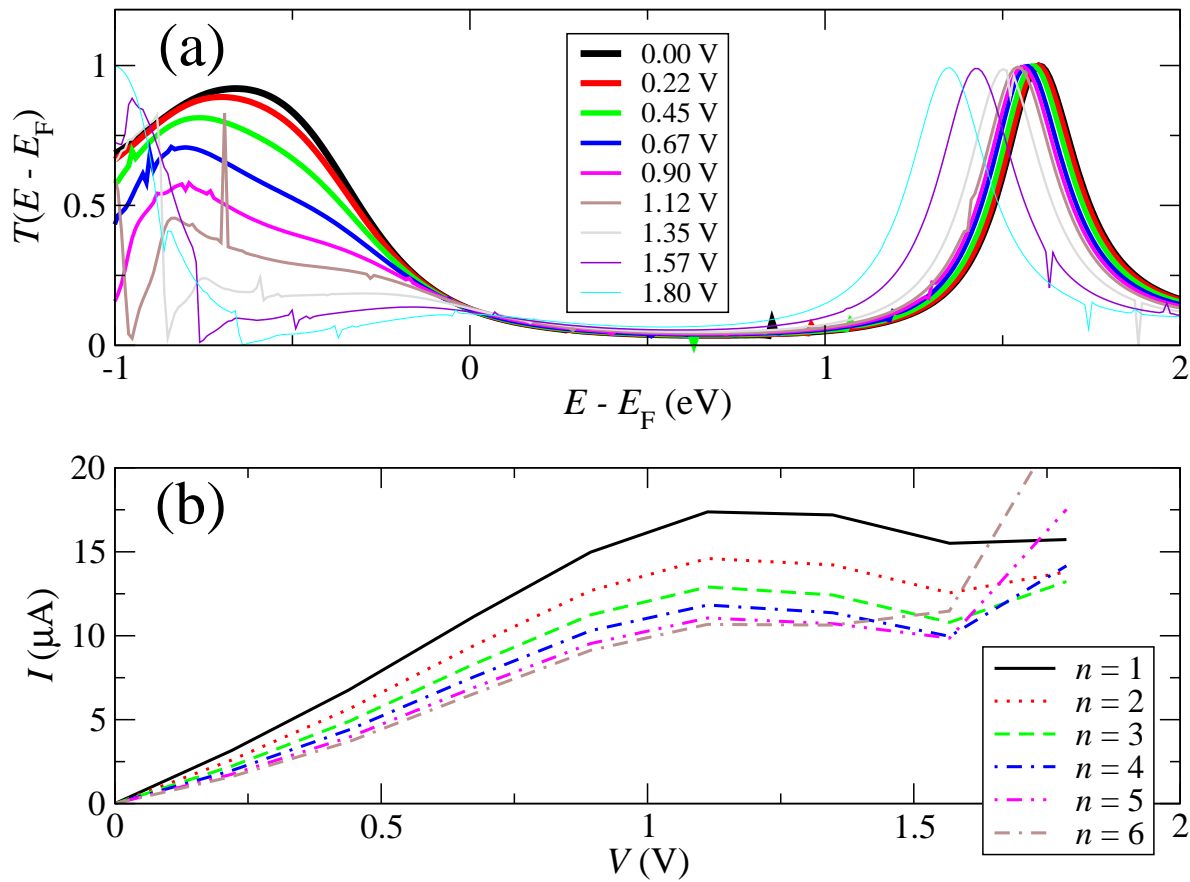}
\caption{\label{Fig2} Transport properties of B-$n$-B molecules.
(a) Transmission coefficients as a function of the bias potential
for $n=3$. (b) $I$-$V$ characteristics as a function of the number
of carbon pairs ($n$) in the chain.}
\end{figure}

The transmission coefficients as a function of the bias voltage
produced by the molecule with thiol connections for $n=3$ carbon
pairs (i.e. 6 atoms) are shown in Fig. (\ref{Fig2}) (a). The
molecule is connected to the three-atoms hollow site, which gives
a strong bond to the surface and therefore produces broad
resonances and high conductances at the Fermi level. At zero bias
the transmission coefficients show the typical two-peak structure
associated with the HOMO and LUMO levels. However, compared to the
case of pure polyynes capped by sulphur atoms \cite{Crl07} these
peaks have a maximum value of only unity. This difference is due
to the fact that only a single channel at the Fermi level is
sustained by the aromatic rings compared with the carbon chain
which has two. These channels come from the $\pi$ orbitals, which
are weakly bonded states and therefore have an energy much closer
to the Fermi level than the strongly bonded $\sigma$ states. The
number of available $\pi$ states is two on the polyyne because on
each carbon atom the hybridization is $sp$ and only two sigma
bonds couple directly to other carbon atoms, leaving two
weakly-bonded $\pi$ orbitals. In the aromatic rings, however, the
hybridization is $sp^2$ and the third sigma bond couples to the
hydrogen atom, leaving only one $\pi$ state on the ring. The
highest electrical resistance is therefore associated with the
rings.

When a bias is applied the most noticeable effect is the different
behavior of the HOMO and LUMO resonances. While the HOMO resonance
is destroyed as the absolute value of the bias increases, the LUMO
resonance remains the same. This different behavior can be
understood by looking at the spatial distribution of the
associated levels. The state related to the HOMO resonance is
largely localized on the polyyne chain and the sulphur atoms,
while it has little weight on the rings. Since these three regions
are only weakly coupled the outer parts move differently under an
applied bias, i.e. one to lower energies and the other to higher
energies, and their electronic occupations change. This can be
clearly seen if an electric field is applied to the isolated
molecule with hydrogens attached to the sulphurs. The spatial
projection of the HOMO shows that the weight on one of the
sulphurs increases while the weight on the other sulphur
decreases. The state becomes therefore asymmetric and, in
accordance with the Breit-Wigner model of resonances \cite{Bre36},
the transmission peak is reduced, as seen in Fig. (\ref{Fig2}) (a)
\cite{TransPol}. The case of the LUMO is however rather different
because this state is mainly localized on the polyyne chain and is
not modified by the bias voltage. Therefore the shape of the LUMO
resonance does not change under bias and the tunneling probability
through it remains the same, as can be seen in Fig. (\ref{Fig2})
(a).

The shift of the HOMO and LUMO resonances with increasing bias is
produced by charge transfer to the leads, which moves the states
to lower energies \cite{Sta06}. This charge transfer is due to
asymmetric partial occupation of states within the bias window
near the HOMO or LUMO resonances. If the Fermi level is close to
the HOMO, like in this case, the weight inside the bias window is
larger for states that were before completely occupied than for
states that were before empty. Therefore the partial occupations
do not sum up to the previous value and the molecule looses
charge. If the Fermi level is however close to the LUMO the effect
is the opposite and the molecule wins charge, as we will see
later.

Fig. (\ref{Fig4}) (a) shows that the zero bias conductance
decreases by less than 2 as $n$ increases from 1 to 6. This trend,
which is similar to what was found in pure polyynes \cite{Crl07}
and agrees with recent preliminary experiments \cite{RNichols}, is
a significant result because it proves it is possible to have
molecular wires that can connect different parts of a circuit with
almost the same conductance.

The presence of NDR in these systems, which is shown in Fig.
(\ref{Fig2}) (b), arises from the sensitivity of the HOMO
resonance to the applied bias. Since the current is calculated by
integrating the transmission coefficient over the bias window,
$I(V)=\frac{2e}{h}\int_{-V/2}^{V/2}\mathrm{d}E\,T(E,V)$, the
suppression of the HOMO resonance can cause the current to
decrease with bias because the reduction of the transmission
compensates the increase of the bias window. For higher voltages
the window reaches again regions with larger conductances and the
current grows again. The shape of the NDR and the location of its
maximum are very similar in all cases, which is a consequence of
the analogous behavior of all these molecules under bias.

\begin{figure}
\includegraphics[width=\columnwidth]{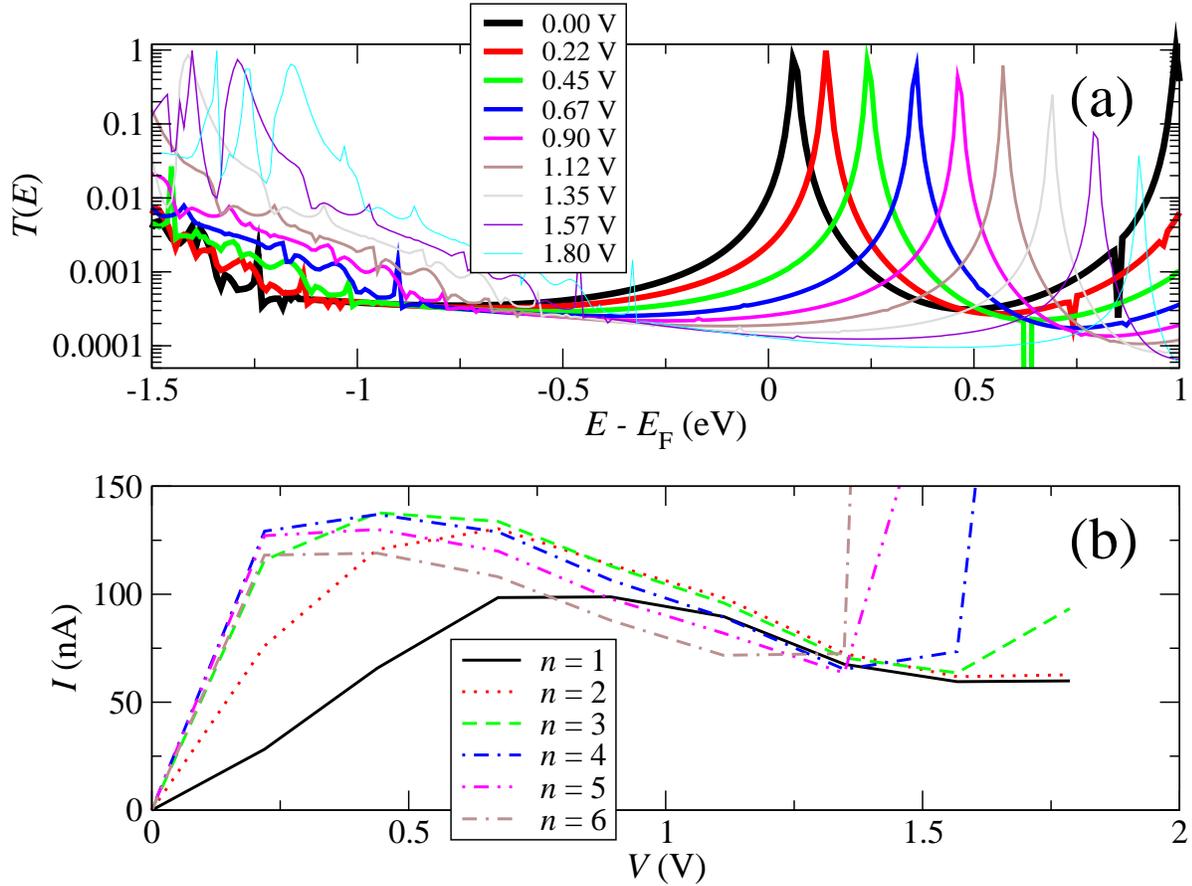}
\caption{\label{Fig3} Transport properties of of P-$n$-P
molecules. (a) Transmission coefficients as a function of the bias
potential. (b) $I$-$V$ characteristics as a function of the number
of carbon pairs ($n$) in the chain.}
\end{figure}

When the benzene-thiol rings are substituted by pyridines, the
presence of nitrogen on the rings produces a series of effects
which greatly modify the behavior of the whole system. On the one
hand, the higher electronegativity of nitrogen compared to that of
sulphur and carbon moves all molecular states downwards,
positioning the Fermi level closer to the LUMO. On the other hand,
the most stable bonding configuration to the surface moves from
the hollow to the top position, which dramatically reduces the
coupling to the electrodes. This reduction can be clearly seen in
the transmission coefficients of Fig. (\ref{Fig3}) (a), where the
resonances are much sharper than in the B-$n$-B case.

The length dependence is even more interesting than in the B-$n$-B
case. As can be seen in Fig. (\ref{Fig4}) (b), the zero bias
conductance increases now with $n$, which is a surprising result.
This behavior, which is due to the movement of the LUMO to lower
energies as the length increases, is not seen in recent
experiments \cite{RNichols}, where a smooth decrease in
conductance with $n$ was observed. The predicted length dependence
is however sensitive to the relative position of $E_\mathrm{F}$,
whose calculated value may be incorrectly predicted by DFT
\cite{Toh07} and experimentally can be affected by the ambient
environment \cite{Che06,Lon06}. Fig. (\ref{Fig4}) (c) shows that
decreasing $E_\mathrm{F}$ by 0.4 eV \cite{FermiPyr}, causes the
conductance to decrease, rather than increase with $n$. This
striking change of tendency can be explained by taking into
account two competing effects. On the one hand, the increasing
distance between the leads reduces exponentially the transmission
in the HL gap, which dramatically reduces the conductance. On the
other hand, the increasing electronic delocalization in longer
molecules decreases the level repulsion and moves the states
downwards, which increases the conductance near the LUMO. As a
consequence, depending on wether the Fermi level is closer or not
to the LUMO one effect dominates the other and the conductance
tendency can change. This result proves that the length dependence
of these molecular wires can be substantially modified and even
reversed by a gate voltage.

As the bias increases, the LUMO resonance in Fig. (\ref{Fig3}) (a)
moves to higher energies, in contrast to Fig. (\ref{Fig2}) (a).
This trend can again be explained by charge transfer under bias,
which in this case moves back to the molecule. The reason now is
due to the location of the Fermi level near the LUMO. The weight
inside the bias window of states that were previously unoccupied
is bigger than the weight of states that were completely occupied
and the molecule wins charge, which moves the states to higher
energies. This movement is the main responsible of the NDR effect.
As the bias window increases it encounters first the LUMO
resonance but at some bias this state rises faster than the bias
window and the current decreases. Notice also the NDR is much more
pronounced in this case compared to the B-$n$-B system and the
positions of the maxima also depend on voltage. The position of
the minimum and subsequent rise in the $I$-$V$ characteristics
moves to lower voltages with increasing $n$, reflecting the
decrease in the HL gap with increasing length. The NDR however
disappears on the same range of bias voltages if the Fermi level
is moved to lower energies, as can be seen in Fig. (\ref{Fig4})
where the current is calculated by integrating the bias-dependent
transmission coefficients on a bias window centered at
$E_\mathrm{F}-0.4$ eV. The continuous increase in the current is
produced now by the proximity of the HOMO resonance which moves to
higher energies as the bias increases. By gating these molecules
it is therefore possible to alter significantly their $I$-$V$
characteristics and tune the appearance of NDR.

\begin{figure}
\includegraphics[width=\columnwidth]{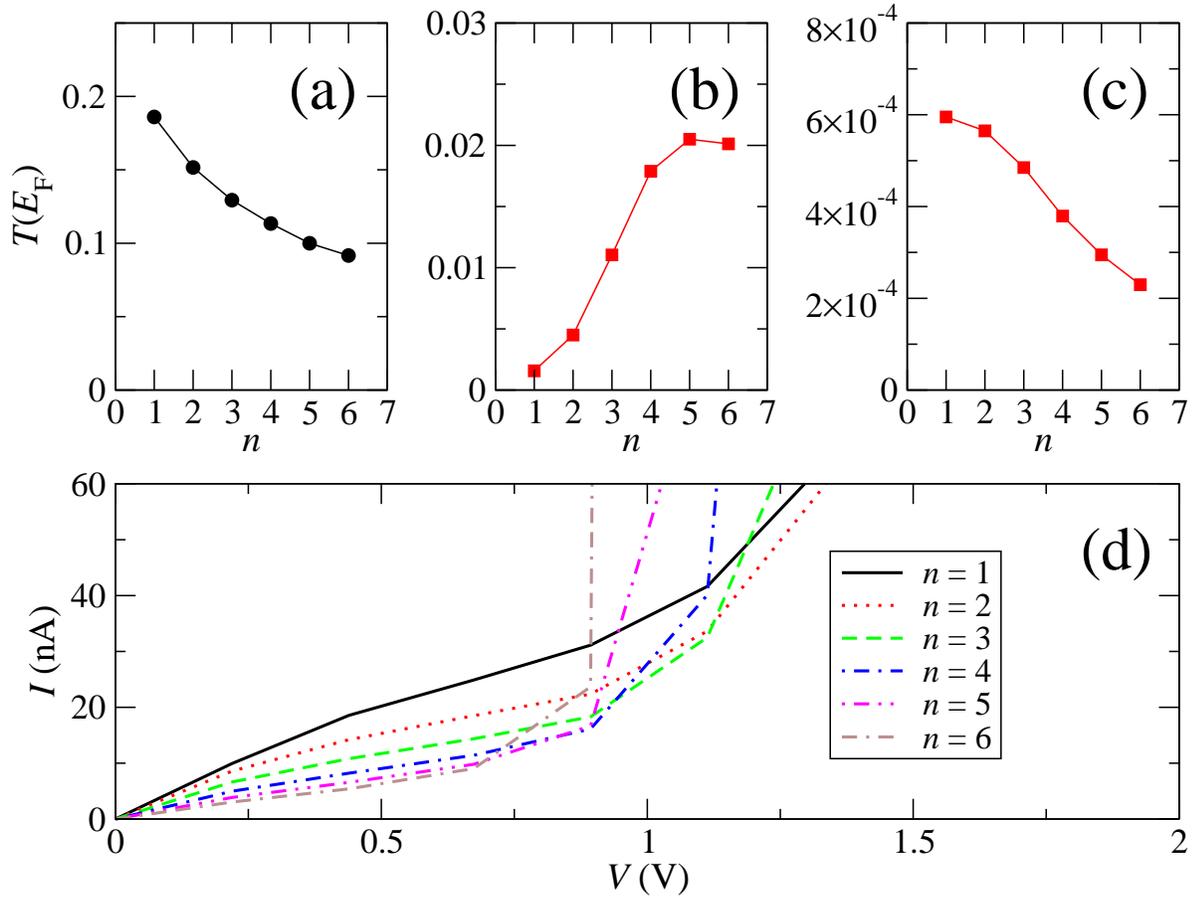}
\caption{\label{Fig4} Zero-bias conductance of (a) B-$n$-B and (b)
P-$n$-P molecules calculated with the Fermi level provided by the
calculation. (c) Zero-bias conductance and (d) $I$-$V$
characteristics for P-$n$-P molecules calculated at
$E_\mathrm{F}-0.4$ eV.}
\end{figure}

In conclusion, we have studied the electronic and transport
characteristics of two novel types of molecules, B-$n$-B and
P-$n$-P, using a combination of density functional theory and
non-equilibrium Green's functions formalism. We found very
interesting properties which make these systems rather promising
from a technological and fundamental point of view. The most
important result is an almost constant length dependence of the
conductance, which is very smooth and under some circumstances can
even increase with the number of carbon pairs in the chain. This
allowed us to predict that it is possible to alter dramatically
the length dependence of molecular wires by using a gate voltage.
We also found the presence of two types of negative differential
resistance. In the B-$n$-B molecule the NDR is due to the
destruction of the HOMO resonance under bias, while in the P-$n$-P
the movement of the LUMO is the main responsible behind the
decrease of the current. We also found that the shape of the NDR
was almost constant in the first molecule while in the second it
depended on the molecular length. Finally, we showed that by
gating the P-$n$-P molecules it is possible to alter the shape of
the $I$-$V$ characteristics quite significantly and make the NDR
disappear for the same range of bias voltages.

\begin{ack}
We thank Martin Bryce, Richard Nichols, Wolfgang Haiss, Simon
Higgins and Santiago Mart\'{\i}n for useful discussions. This work
was supported by the European Comission, Qinetiq, and the British
EPSRC, Department of Trade and Industry, Royal Society and
Northwest Regional Development Agency.
\end{ack}


\begin{thebibliography}{99}
\bibitem{LeadsCalc}
Since both leads were the same only one computation was enough to
calculate the electronic properties of the electrodes.
\bibitem{TransPol}
Note this effect is not found in pure polyynes \cite{Crl07}
because the HOMO is equally distribuyed across the molecule and
remains the same under bias.
\bibitem{FermiPyr}
One should be cautious here because when the Fermi level is moved
the charge on the molecule changes and therefore the Hamiltonian
and by extension the transmission coefficents are modified.
However, if $E_\mathrm{F}$ remains inside the HL gap the overall
charge changes are negligible and the mean-field Hamiltonian
remains the same.
\end{thebibliography}
\end{document}